\documentclass[12pt,letterpaper]{article}

\usepackage[activeacute,english]{babel}
\usepackage{amsmath}
\usepackage{amsbsy}
\usepackage{amsthm}
\usepackage{amssymb}
\usepackage{latexsym}
\usepackage[dvips]{graphicx}

\begin{document}

\title{\textbf{Baryonic resonances mass spectrum from a modified
perturbative QCD}}

\date{}
\author{}
\maketitle

\begin{center}
\vspace{-1.0cm}{\large \textbf{Alejandro Cabo Montes de
Oca}}\\[0pt] \vspace{.25cm} {\small \textit{Instituto de
Cibern\'{e}tica, Matem\'{a}tica y F\'{\i}sica; }}\\[0pt] {\small
\textit{Calle E, No. 309, Vedado, La Habana, Cuba,}}\\[0pt]
\vspace{.5cm}{\large \textbf{Marcos Rigol Madrazo,}} \\[0pt]
\vspace{.25cm} {\small \textit{Centro de Estudios Aplicados al
Desarrollo Nuclear}} \\[0pt] {\small \textit{Calle 30, N. 502 e/
5ta y 7ma, Miramar, La Habana, Cuba }}
\\[0pt]
\end{center}

\begin{abstract}
A recently proposed modified perturbation expansion for QCD is
employed to evaluate the quark self-energies. Results of the order
of 1/3 of the nucleon mass are obtained for the effective masses
of the up and down quarks in a first approximation. Also, the
predicted flavor dependence of the calculated quarks masses turns
out to be the appropriate to well reproduce the spectrum of the
ground states within the various groups of hadronic resonances
through the simple addition of the evaluated constituent quark
masses. The results suggests to conjecture that the modified
expansion, after also introducing quark condensates in a same
token as the gluonic ones, would be able to furnish a natural
explanation of the mass spectrum of the three generations of
fundamental fermions.
\end{abstract}

\newpage

Strong interactions at low energy are the subject of an intense
research activity directed to complete the present understanding
of QCD \cite {ShuryakTex,Narison}. Two of the many important
questions to be clarified are the nature of the mass spectrum for
the numerous hadron and meson resonances and a better
understanding of the differences between the constituent and
Lagrangian quark masses \cite{weinberg1,weinberg2,weinberg3}. On a
more general level, there is also a little understanding of the
physics determining the mass spectrum of the three generations of
fundamental fermions \cite{weinberg1}.

In previous papers we have proposed a modification of the standard
perturbative expansion for QCD which is based in the idea that the
low energy properties are determined by the presence of a
condensate of gluon pairs \cite{1995,prd,tesis}. In paper
\cite{1995} the general structure of the modified Feynman rules
were proposed and showed that it produces a non-vanishing value of
$\langle G^2\rangle$ even in the simplest approximation and that
the condensate seems to be generated spontaneously from the
perturbative vacuum of zero particles in a similar form as the
earlier Savvidy chromomagnetic vacuum field. Further, in paper
\cite{prd,tesis} the state of the non-interacting theory which
shows the proposed kind of modified Feynman rules, was determined.
It occurred that the state is constructed as a coherent
superposition of zero momentum gluon pairs. For its determination
the condition of being a physical state (a zero mode of the BRST
charge) of the non-interacting theory was imposed. After that, the
operational quantization was considered in a scheme of the
Gupta-Bleuler type in which all the gluons are considered on the
same footing \cite{kugo}. This corresponds with the selection
$\alpha =1$ value of the gauge parameter in the standard
functional approach. In justifying to proceed in such a way, our
central assumption is that under the adiabatic connection of the
color interaction, the evolution will not bring the state out of
the physical subspace at any stage of the connection. Thus, the
final state will be also a physical state of the interacting
theory in which the associated Feynman expansion should have a
physical meaning. It is clear that this approach leaves out the
question of the construction of a gauge parameter independent
formulation of the theory. However, as a minimal logical ground
for the physical relevance of the predictions is given, we decided
to postpone this more technical question to further analysis to be
done.

Therefore, in the present work the proposed expansion is applied
to evaluate the mass spectrum of the quarks as modified by the
inclusion of the gluon condensation effects. The main result which
follows is that a particular branch of the spectrum produces
constituent mass values of the order 1/3 of the nucleon mass for
the up and down quarks. Further, the result for flavor dependence
of the effective quark masses is able to predict the rest masses
of many of the ground state resonances in the various groups of
them as classified in \cite{Report} after a simple addition of the
values associated to the corresponding constituent quarks. It
should be mentioned that the constituent quark predictions of the
order of 1/3 of the nucleon mass as determined by gluon
condensation effects were also obtained in \cite{Steele1,Steele2}
by employing a different approach. Furthermore, the spectrum of
mesons, which are currently accepted to show constituent quarks
structure are also predicted. In our view, these results strongly
support the important role of gluon condensation in determining
the structure of many hadrons and mesons and also the possibility
of describing them with a modified perturbative expansion. If such
is the case, various non-perturbative characteristics of the QCD
could be treated in a similar way as in the theory of Bose
Condensation \cite{abrikosov}. The non-perturbative properties of
the theory, then, would be described by the condensation parameter
controlling the modified expansion. The possibility for such a
picture was already suggested by the earlier chromomagnetic field
models \cite{Savvidy1,Savvidy2,Savvidy3}. It is natural to expect
that the Lorentz invariance of the ground state and the vector
character of the gluonic field makes it natural that a Bose
condensate effect for gluonic field should present peculiar
characteristics as in the case of superconductivity
\cite{abrikosov}. However, this situation is not excluding the
possible applicability of a modified perturbative expansion as in
standard Bose condensed systems \cite{abrikosov}. The results for
the mass spectra arising from the calculations shown here support
the existence of such a picture.

The alternative initial state determining the modified
perturbation expansion through the Wick theorem was found in
\cite{prd,tesis} to have the form
\begin{equation}
\mid \phi \rangle =\exp \sum\limits_a\left( \frac
12A_{0,1}^{a+}A_{0,1}^{a+}+\frac
12A_{0,2}^{a+}A_{0,2}^{a+}+B_0^{a+}A_0^{L,a+}+
i\overline{c}_0^{a+}c_0^{a+}\right) \mid 0\rangle \label{Vacuum1}
\end{equation}
in terms of the transverse and longitudinal gluon $A_{0,1}^{a+},
A_{0,2}^{a+},A_0^{L,a+}$, ghosts $c_0^{a+},\overline{c}_0^{a+} $,
and Nakanishi-Lautrup $B_0^{a+}$ creation operators of the
interaction free theory \cite{kugo}. The state is colorless as
indicated by the contracted color index $a$. For its construction
the BRST physical state conditions $Q_B\mid \Phi \rangle =0$ and
$Q_C\mid \Phi \rangle =0$ were required, in which $Q_B$ and $Q_C$
are the charges associated to the BRST symmetry and ghost number
conservation.

In Refs. \cite{prd,tesis} the state (\ref{Vacuum1}) was sought in
order to implement that the net effect of the application of the
Wick Theorem to the expansion of the evolution operator be to
produce the form of the gluon propagator introduced in
\cite{1995}. The expression of the modified propagator in the
momentum representation is
\begin{equation}
G_{0\mu\nu}^{ab}\left( k\right)=\delta ^{ab}g_{\mu\nu}\left[ \frac
1{k^2}-iC\delta \left( k\right) \right], \label{prop}
\end{equation}
in which $C$ is the parameter associated with the gluon
condensate, and was determined to be real and nonnegative
\cite{prd,tesis}. The ghost and fermion propagators do not showed
a modification. It will be the case that under assuming the
existence of fermion condensates such a modification should
appear. Precisely this possibility leads to the conjecture to be
posed in the concluding remarks.

In order to fix a physically supported value for the parameter
$C$, the following recourse was employed \cite{1995}. The gluon
condensate parameter $\langle G^2\rangle$ was calculated up to the
order $g^2$ and the resulting function of the parameter $C$ used
to determine this constant by selecting its value to give the
presently accepted estimate of $\langle g^2G^2\rangle$ in the
physical vacuum. The details of this evaluation are presented in
an extended version of this article \cite{extend}. For the
calculation of $\langle g^2G^2\rangle$ the following expression
was employed
\[
\left\langle 0\right| S_g\left| 0\right\rangle =\left[ \frac
1N\int D\phi S_g\left[ \phi \right] \exp \left( S_T\left( \phi
\right) \right) \right], \label{sg}
\]
where $S_g\left[ \phi \right]$ represents the usual gauge
invariant gluon part of the action and $S_T$ is the total
Fadeev-Popov action for $\alpha =1$.

The final result for $\langle g^2G^2\rangle$ in terms of $C$ is
\[
\langle g^2G^{2}\rangle =\frac{288g^{4}C^{2}}{\left(2\pi \right)
^{8}}.
\]
Further, by making use of the estimated value for $\langle
g^2G^2\rangle \cong 0.5\left( GeV/c^2\right)^4$ \cite{ShuryakTex},
is obtained the result $g^2C=64.9394\ \left(GeV/c^2\right)^2$.

After the phenomenological determination of $g^2C$, let us
consider in what follows the one loop correction to the quark
self-energy. The notation employed is the one in Ref. \cite{Muta}.
The inverse quark propagator has the form
\begin{equation}
G_{2ij}^{-1}\left( p\right) =i\delta _{ij}\left( m_Q-p^\mu \gamma
_\mu -\Sigma \left( p\right) \right), \label{Green1}
\end{equation}
in which the self-energy part $\Sigma \left( p\right)$ is
determined up to the order $g^2$ and $m_Q$ is the Lagrangian mass
of the specific kind of flavor being considered.

The only change appearing in (\ref{Green1}) with respect to the
similar calculation in the standard perturbation expansion is
related with the gluon propagator $G_{0\mu\nu}^{ab}\left(
k\right)$ to be used which will include the mentioned condensate
term as defined in (\ref{prop}). Here we will be concerned only
with the evaluation of the contribution to the self-energy of the
condensate term. The aim is to investigate within the most simple
approximation the predictions for the mass corrections determined
by it. Therefore, after integrating in the expression for $\Sigma
\left( p\right)$ it follows for the inverse Green function
(\ref{Green1})
\begin{equation}
G_{2ij}^{-1}\left( p\right) =i\delta _{ij}\left( m_Q\left(
1+2\frac{M^2}{ \left( m_Q^2-p^2\right) }\right) -p^\mu \gamma _\mu
\left( 1+\frac{M^2}{ \left( m_Q^2-p^2\right) }\right) \right),
\label{Green2}
\end{equation}
where $M^2=0.111111\ \left(GeV/c^2\right)^2$.

The vanishing of the determinant of the matrix (\ref{Green2})
determines the modified mass shell in the considered approximation
including the effects of the condensate on the spectrum. The mass
values will depend on the quark flavor through their Lagrangian
mass parameters $m_Q$ which have been determined by independent
means \cite{weinberg1}.

After solving the cubic equation produced by equating to zero the
determinant of (\ref{Green2}), the following quark mass shells
relations are obtained
\[
p^2-m_{q,l}^2=0;\ \ \text{for} \;l=1,2,3
\]
in which $m_{q,l}^2,l=1,2,3$ designates the three different
analytic solutions for the squared masses written below:
\begin{eqnarray*}
m_{q,1}^2 &=&A^{\frac 13}-B+m_Q^2+\frac 23M^2, \\ m_{q,2}^2
&=&-\frac 12A^{\frac 13}+\frac 12B+m_Q^2+\frac 23M^2+\frac
12i\sqrt{3} \left( A^{\frac 13}+B\right), \\ m_{q,3}^2 &=&-\frac
12A^{\frac 13}+\frac 12B+m_Q^2+\frac 23M^2-\frac 12i\sqrt{3}
\left( A^{\frac 13}+B\right),
\end{eqnarray*}
where the quantities $A$, $B$ are determined by the expressions
\begin{eqnarray*}
A&=&\frac{5}{6}m^{2}_{Q}M^{4}-\frac{1}{27}M^{6}+\frac{1}{18}\left(
96m^{6}_{Q}M^{6}+177m^{4}_{Q}M^{8}-12m^{2}_{Q}M^{10}\right)^{\frac{1}{2}},
\\ B
&=&\frac{\frac{2}{3}m^{2}_{Q}M^{2}-\frac{1}{9}M^{4}}{A^{\frac{1}{3}}}.
\end{eqnarray*}

\begin{figure}[h]
\begin{center}
\includegraphics[scale=0.44,angle=-90]{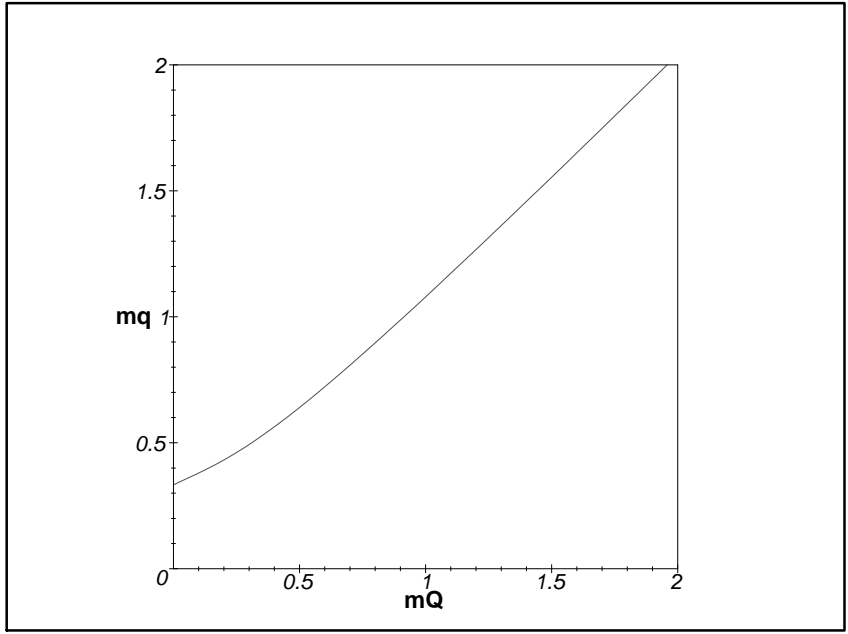}
\end{center}
\caption{Real solution for the quark mass as a function of the
Lagrangian mass (masses in units of $GeV/c^2$).} \label{Mass1}
\end{figure}

The various dispersion relations arising from these solutions have
some degree of redundancy. However, only six independent solutions
emerge at the end for all of the $m_{q,l}^2$, with $l=1,2,3$. The
most of these six branches lead to imaginary values of the masses
in some regions of the $m_Q$ values. The physical nature of these
fermionic wave modes deserve a closer examination which will be
considered elsewhere. Here we limit ourselves to examine one of
the solutions for the squared mass parameter which shows positive
values $m_q$ for the fermion squared masses at any momenta and
also a growing dependence of the mass on the Lagrangian mass
parameter $m_Q$. The plot of the graph $m_q$ as function of $m_Q$
for this solution is shown in Fig.\ref{Mass1} in the region
$m_Q<2\ GeV/c^2$.

As it can be observed from Fig.\ref{Mass1}, the interaction with
the vacuum condensate dress an initially assumed massless quark
$(m_Q=0)$ by giving it a finite mass value of the order of one
third of the nucleon mass. That is, the $u$ and $d$ $quarks$
acquire in this propagation mode a high contribution to their
masses due to the interaction with the condensate. In Table
\ref{table1} the mass values of the considered propagation mode
for each of the six quark flavors as characterized by their
reported Lagrangian mass values $m_Q$ in Report \cite{Report} are
shown.

\begin{table}[h]
\caption{Quark mass values in presence of the condensate in units
of $MeV/c^2$.}\label{table1}
\begin{center} \begin{tabular}{||c||c||c||c||}
\hline\hline $Quarks$ & $m_{Low}^{Exp}$ & $m_{Up}^{Exp}$ &
$m_{Med}^{Theo}$
 \\ \hline\hline (u) & 1.5 & 5 & 334.944
 \\ \hline\hline (d) & 3 & 9 & 336.287
 \\ \hline\hline (s) & 60 & 170 & 388.191
 \\ \hline\hline  (c) & 1100 & 1400 & 1315.241
 \\ \hline\hline (b) & 4100 & 4400 & 4269.572
 \\ \hline\hline (t) & 168600 & 179000 & 173800.48
 \\ \hline\hline
\end{tabular}\end{center}
\end{table}
\noindent where $m_{Low}^{Exp}$ is the reported lower bound value
for the Lagrangian mass, $m_{Up}^{Exp}$ is the reported upper
value for the Lagrangian mass and $m_{Med}^{Theo}$ is the
calculated mean value of the constituent mass, for the lower and
upper bound of the Lagrangian masses.

Table \ref{table1} shows that the masses of the $u$ and $d$ quarks
are near one third of the nucleon mass. It can be noticed that the
$c,b$ and $t$ quarks show masses for which the effect of gluon
condensation is not neatly evidenced. The light quarks $u,d$ and
$s$ are the ones which get a clear influence on their mass due to
gluon condensation as described here.

These results support that the expansion introduced in Ref.
\cite{1995,prd,tesis} could have the chance of furnishing
important non-perturbative information on the QCD low energy
physics. This conclusion is further strengthened by evaluating the
masses of the ground states within the various groups of baryonic
resonances as classified in \cite{Report}. The calculation is done
by simply adding the masses corresponding to each of the types of
quarks constituting the specific hadron being examined. Table
\ref{table2} shows the results of that evaluation for the lower
energy states within the groups of baryon resonances in
conjunction with their experimentally determined masses as given
in \cite{Report}.

\newpage

\begin{table}[h]
\caption{Experimental and Theoretical Baryonic Resonance Masses in
units of $MeV/c^2$.} \label{table2}
\begin{center}\begin{tabular}{||r||r||r||r||} \hline\hline Baryon &
Exp.Val. & Th.Mean.Val. & Rel.Err. \\ \hline\hline p(uud) \ \  &
938.27231 & 1006.175 & 7.24 \\ \hline\hline n(udd) \ \  &
939.56563 & 1007.519 & 7.23 \\ \hline\hline $\Lambda $(uds) \ \  &
1115.683 & 1059.422 & 5.04 \\ \hline\hline $\Sigma ^{+}$(uus)\ \ \
& 1189.37 & 1058.078 & 11.04
\\ \hline\hline $\Sigma ^{0}$(uds) \ \  & 1192.642 & 1059.422 &
11.17 \\ \hline\hline $\Sigma ^{-}$(dds)\ \ \  & 1197.449 &
1060.766 & 11.41 \\ \hline\hline $\Xi ^{0}$(uss)\ \ \  & 1314.9 &
1111.325 & 15.48 \\ \hline\hline $\Xi ^{-}$(dss) \ \  & 1321.32 &
1112.669 & 15.79 \\ \hline\hline $\Omega ^{-}$(sss)\ \ \  &
1642.45 & 1164.572 & 29.10 \\ \hline\hline $\Lambda
_{c}^{+}$(udc)\ \ \  & 2284.9 & 1986.472 & 13.07 \\ \hline\hline
$\Xi _{c}^{+}$(usc) \ \  & 2465.6 & 2038.375 & 17.33 \\
\hline\hline $\Xi _{c}^{0}$(dsc)\ \ \  & 2470.3 & 2039.719 & 17.43
\\ \hline\hline $\Omega _{c}^{0}$(ssc)\ \ \  & 2704 & 2091.622 &
22.65 \\ \hline\hline $\Lambda _{b}^{0}$(udb)\ \ \  & 5624 &
4940.803 & 12.15 \\ \hline\hline
\end{tabular}\end{center}
\end{table}

As it can be noticed, the values in Table \ref{table2} reasonably
well match the mass spectrum of the ground states in each of the
families of baryons as classified in \cite{Report}. Let us comment
about some possible sources of theoretical errors associated to
the evaluation of the quark masses and the baryon spectrum. We
estimate that one of them is the non inclusion of the standard one
loop self energy contributions for the quarks. It should be
stressed that its proper consideration within the momentum scale:
$p<1$ GeV, needs for a knowledge of the running coupling in this
region. However, the influence of the modified propagator on the
coupling reflects itself only through a two loop evaluation of
this quantity. This is because, the one loop divergences are
unchanged by the use of the new propagator. In connection with the
relative influence of the appreciable lack of precision in the
current masses it could be observed that: a) For the mainly
massless u and d quarks, the results obtained can not be greatly
influenced by the big errors in the estimates of current masses,
b) For high values of the current masses for c, b, and t quarks
the influence of the condensate can be expected to be weak and the
prediction should coincide with the current masses and c) On the
other hand for the strange s quark, the effect of the condensate
is of the same order as the current mass value. Thus, the error in
the estimate of this magnitude could become relevant for the
evaluation of the effective mass. This argument, might be working
for the results for the baryons in Table II including u and d
quarks plus one, two or three s quarks. An increase in the
effective s quark mass of near 180 $MeV/c^2$ seems to very much
improve the predictions for such resonances. Another relevant
source of errors is the high margin of error of the present
estimates for the current masses. Clearly, the interaction effects
could have their influence in the results. Their consideration,
however, requires a further study of the predictions for the bound
state spectrum of quarks.

Next, the calculated masses for a group of vector mesons is
depicted in Table \ref{table3} in conjunction with their reported
experimental values \cite{Report}. The evaluation is performed
again by simply adding the mass of the corresponding quarks
entering in the known composition of each meson.

\begin{table}[h]
\caption{Experimental and Theoretical Masses for a group of Vector
Mesons in units of $MeV/c^2$.}\label{table3} 
\begin{center}\begin{tabular}{||r||r||r||r||} \hline\hline Meson & Exp.Val.
& Th.Mean.Val. & Rel.Err.\\ \hline\hline $\rho \left(
\frac{u\overline{u}-d\overline{d}}{\sqrt{2}}\right) $ & 770.0 &
671.231 & 12.83 \\ \hline\hline $\varpi \left(
\frac{u\overline{u}+d\overline{d}}{\sqrt{2}}\right) $ & 781.94 &
671.231 & 14.16 \\ \hline\hline $\phi \left( s\overline{s}\right)
$ \ \ \ \  & 1019.413 & 776.381 & 23.84 \\ \hline\hline $J/\psi
\left( 1S\right) \left( c\overline{c}\right) $ & 3096.88 &
2630.482 & 15.06 \\ \hline\hline $Y\left( 1S\right) \left(
b\overline{b}\right) $ & 9460.37 & 8539.144 & 9.74 \\ \hline\hline
\end{tabular}\end{center}
\end{table}

As it can be observed from the results for the vector mesons and
baryonic resonance masses, in spite of the simple approximation
which has been considered, the general structure of the mass
spectrum is well reproduced under the only assumption of the
accepted value of the gluon condensation parameter $\langle g^2G^2
\rangle$, \cite{ShuryakTex}.

It should underlined that among the obtained dispersion relations,
there is a solution predicting a vanishing value of the mass when
$m_Q\rightarrow0$. It might be the case that these modes could be
connected with the family of low mass mesons (e.g. $\Pi^{\pm,0}$
mesons). Specifically, the possibility exists that the bound
states of quark excitations in these low mass states of quarks
could describe such low lying mesons. An analysis of this question
however, will be deferred to a future extension of the work. As a
last point, we would like to conjecture on a possibility suggested
by the analysis given here. It is related with the question about
whether the mass spectrum of the whole three generations of
fundamental fermions could be predicted by a slight generalization
of the modified perturbation theory under consideration. In this
sense, the presented results led us to the idea that after the
introduction of quark condensates along the same lines as it was
done for gluon ones in \cite{1995,prd,tesis}, the obtained
perturbation expansion can have the chance to predict both, the
Lagrangian mass and the constituent quark mass spectra of the
three families of fundamental fermions. The fermion condensates as
described in the proposed perturbative way, would be encharged to
produce the Lagrangian quark masses as usual, through the chiral
symmetry breaking. The gluonic condensates, in one hand, and as
illustrated here, could be responsible of generating states of
large constituent mass for the low mass quarks states. In another
hand, it seems feasible that the higher order radiative
corrections (including color interactions with the condensate as
well as chiral symmetry corrections) could also determine the mass
spectra for leptons and neutrinos. The smaller scale for the
masses of these particles could be produced by the lack of lower
order color interaction terms in their self-energy. Therefore, the
possibility that the Lagrangian mass spectrum of the three
generations of the fundamental fermions could be predicted by a
modified perturbation expansion of the sort being proposed is
suggested. Work devoted to investigate the above conjecture will
be considered elsewhere.

\textbf{Acknowledgments}

The authors wish to deeply acknowledge the helpful advice and
comments of Profs. J. Pestieau, J. Alfaro, M. Lowe, C. Friedli and
M. Hirsch. One of the authors (A.C.M) also would like to express
his gratitude to the Abdus Salam ICTP and in particular to its
Associate-Membership Programme for the whole general support.

\end{document}